\documentclass[8.5pt,jcp,twoside,twocolumn]{article}
\usepackage[super,sort&compress,comma]{natbib}
\usepackage[version=4]{mhchem}
\usepackage{times,mathptmx}
\usepackage{sectsty}
\usepackage{balance}
\usepackage{xspace}
\usepackage{algorithm,algorithmic}

\usepackage{epsfig,amssymb,amsmath,amsthm,amsfonts,amsbsy,mathrsfs}
\usepackage{graphicx,color}
\usepackage{amsmath,eqnarray}
\usepackage{amssymb}
\usepackage{epstopdf}
\usepackage{graphicx} 
\usepackage{lastpage}
\usepackage[format=plain,justification=raggedright,singlelinecheck=false,font=small,labelfont=bf,labelsep=space]{caption}
\usepackage{fancyhdr}
\usepackage{braket}
\usepackage{caption}
\usepackage{url}

\begin{document}

\thispagestyle{plain} \fancypagestyle{plain}{
\renewcommand{\headrulewidth}{1pt}}
\renewcommand{\thefootnote}{\fnsymbol{footnote}}
\renewcommand\footnoterule{\vspace*{1pt}%
\hrule width 3.4in height 0.4pt \vspace*{5pt}}
\setcounter{secnumdepth}{5}

\makeatletter
\def\subsubsection{\@startsection{subsubsection}{3}{10pt}{-1.25ex plus -1ex minus -.1ex}{0ex plus 0ex}{\normalsize\bf}}
\def\paragraph{\@startsection{paragraph}{4}{10pt}{-1.25ex plus -1ex minus -.1ex}{0ex plus 0ex}{\normalsize\textit}}
\renewcommand\@biblabel[1]{#1}
\renewcommand\@makefntext[1]%
{\noindent\makebox[0pt][r]{\@thefnmark\,}#1} \makeatother
\renewcommand{\figurename}{\small{Fig.}~}
\sectionfont{\large}
\subsectionfont{\normalsize}

\fancyfoot{}
\fancyfoot[RO]{\footnotesize{\sffamily{1--\pageref{LastPage}
~\textbar  \hspace{2pt}\thepage}}}
\fancyfoot[LE]{\footnotesize{\sffamily{\thepage~\textbar\hspace{3.45cm}
1--\pageref{LastPage}}}} \fancyhead{}
\renewcommand{\headrulewidth}{1pt}
\renewcommand{\footrulewidth}{1pt}
\setlength{\arrayrulewidth}{1pt} \setlength{\columnsep}{6.5mm}
\setlength\bibsep{1pt}
\newcommand{\vt}{\vec{\theta}}

\twocolumn[
\begin{@twocolumnfalse}

\noindent\LARGE{\textbf{Hybrid Hamiltonian Simulation for Excitation Dynamics}} \vspace{0.6cm}\\
\noindent\large{ Linyun Wan\textit{$^{a}$}, Jie Liu\textit{$^{b\ast}$}, Zhenyu Li\textit{$^{a,b}$} and Jinlong Yang\textit{$^{a,b\ddag}$}} \\

\noindent\textit{\small{\textbf{Received Xth XXXXXXXXXX 20XX, Accepted Xth XXXXXXXXX 20XX\newline First published on the web Xth
XXXXXXXXXX 200X}}}

\noindent \textbf{\small{DOI: 00.0000/00000000}} \vspace{0.6cm}

\noindent \normalsize{
Hamiltonian simulation is one of the most anticipated applications of quantum computing. Quantum circuit depth for implementing Hamiltonian simulation is commonly time dependent using Trotter-Suzuki product formulas so that long time quantum dynamic simulations (QDSs) become impratical on near-term quantum processors. Hamiltonian simulation based on Cartan decomposition (CD) provides an appealing scheme for QDSs with fixed-depth circuits, while it is limited to time-independent Hamiltonian. In this work, we generalize this CD-based Hamiltonian simulation algorithm for studying time-dependent systems by combining it with variational quantum algorithms. The time-dependent and time-independent parts of the Hamiltonian are treated using variational and CD-based Hamiltonian simulation algorithms, respectively. As such, this hybrid Hamiltonian simulation requires only fixed-depth quantum circuits to handle time-dependent cases while still maintaining high accuracy. We apply this new algorithm to study the response of spin and molecular systems to $\delta$-kick electric fields and obtain accurate spectra for these excitation processes.}

\vspace{0.5cm}
\end{@twocolumnfalse}
]

\footnotetext{\textit{$^{a}$Key Laboratory of Precision and Intelligent Chemistry, University of Science and Technology of China,    Hefei, Anhui 230026, China}}
\footnotetext{\textit{$^{b}$Hefei National Laboratory, University of Science and Technology of China, Hefei, Anhui, 230026, China}}  
\footnotetext{\textit{$^{*}$~liujie86@ustc.edu.cn}}
\footnotetext{\textit{$^{\ddag}$~jlyang@ustc.edu.cn}}

\section{INTRODUCTION}
Quantum dynamics simulation (QDS) is widely used to study interactions of light and matter, which result in a variety of interesting photoinduced physical and chemical processes, such as electron transitions, energy and charge transfers, and photocatalysis~\cite{NelWhiBjo20,MieOllTac23}. Simulating these processes by solving the time-dependent Schr\"odinger equation (TDSE) suffers from an exponential scaling of computational cost with respect to the system size, and is therefore limited to toy model systems consisting of few electronic (and nuclear) degrees of freedom. Quantum computing provides a new computational paradigm for efficiently solving dynamic simulation problems for complex quantum systems. This efficiency stems from the same fundamental principles of quantum mechanics that govern quantum computers and QDSs~\cite{feynman2018simulating,lloyd1996universal,wiesner1996simulations,zalka1998efficient,tacchino2020quantum}.

Quantum computing is anticipated to overcome the exponential wall problem by virtual of quantum state superposition and entanglement~\cite{feynman2018simulating}. An appropriate quantum algorithm offers exponential, or at least polynomial, speedup compared to the best classical algorithms, allowing us to solve QDS problems on quantum computers with favorable scaling~\cite{daley2022practical}. Many recent effects have been devoted to developing novel quantum algorithms for simulating time evolution of a many-body fermionic or spin systems~\cite{PhysRevResearchMeasurementLee,smith2019simulating}. Given an initial state $\ket{\Psi(0)}$, the time evolution of the quantum system is governed by the TDSE
\begin{equation}\label{eq:TDSE}
    i\frac{\partial |\Psi(t)\rangle}{\partial t} = H |\Psi(t)\rangle
\end{equation}
and its solution at time $t$ is given as 
\begin{equation}
    |\Psi(t)\rangle = e^{-i H t} |\Psi(0)\rangle,
\end{equation}
with $H$ being the system Hamiltonian. In QDSs, an efficient implementation of the time propagator $e^{-i H t}$, known as Hamiltonian simulation, on quantum computers is at the core of this issue~\cite{georgescu2014quantum}.

Many techniques for Hamiltonian simulation have been developed for general Hamiltonians, such as Trotter-Suzuki algorithm (product formulas)~\cite{berry2007efficient,hatano2005finding}, linear combinations of unitary operations (LCU) algorithm~\cite{childs2012hamiltonian,BerChiCle15}, quantum walk~\cite{lovett2010universal}, Trotter-Suzuki with Lie algebra algorithm~\cite{somma2016trotter}, and quantum singular value transformation~\cite{low2017optimal,kikuchi2023realization}. Here, given a fault-tolerant quantum computer, the Trotter-Suzuki product formulas offers a simple and straightforward approach to carry out Hamiltonian simulation~\cite{lloyd1996universal,ChiSuTra21}. Taylor expansion-based Hamiltonian simulation that approximates the time evolution operator as a linear combination of unitary operators exihibits a better asympotic scaling~\cite{BerChiCle15}. These methods make one believe that, for the Hamiltonian of a real physical system, the corresponding time evolution operator can be executed on a quantum computer using polynomial gates with error correction~\cite{MieOllTac23}. 

Despite ongoing advancements in quantum computing techniques, quantum devices remain susceptible to significant errors resulting from noise in the near future. They are limited to applying a small number of operations within the coherence time on a few qubits, a characteristic of what we term noisy intermediate-scale quantum  (NISQ) computers~\cite{preskill2018quantum}. Consequently, devising quantum algorithms that are composed of shallow circuits and resource-efficient ansatzes is critical for running quantum simulations on near-term quantum computers. 
It is in general able to explore the symmetry of the system~\cite{tran2021faster}, the feature of the initial state~\cite{ward2009preparation}, and the algebraic properties of the system~\cite{camps2022algebraic} to reduce the circuit depth required to execute QDSs. Variational quantum algorithms, including variational quantum dynamics simulation (VQDS) algorithm~\cite{endo2020variational,cerezo2021variational} and adaptive variational quantum dynamics simulation (ADAPT-VQDS) algorithm~\cite{barison2021efficient}, had been also proposed as near-term schemes for running QDSs. These methods maintain a shallow circuit depth during the time evolution process by variationally minimizing the distance between a trial state that is prepared with parameterized circuits and the exact time-evolution state. One of the main limitations of variational approaches arises from the matrix inversion and that is sensitive to noise and condition number. To avoid the matrix inversion, variational method combined with product formulation have been proposed, like variational fast-forward algorithm~\cite{cirstoiu2020variational}, Hardware-efficient variational quantum algorithms~\cite{benedetti2021hardware} and so on. However, another limitation still exist, which arises from the expressivity of the circuit ansatz. This problem is even more pronounced in QDSs compared to static electronic structure simulations~\cite{barison2021efficient}. On the other hand, a small time step is always necessary to guarantee computational accuracy, especially in long-term variational QDSs.

To simulate quantum dynamics with fixed-depth circuits, Cartan decomposition(CD) has been suggested as the state-of-the-art technique for constructing an accurate decomposition of time-evolution operators, regardless of the total simulation time~\cite{KokSteWan22}. Here, one first builds the Lie algebra generated by the time-independent Hamiltonian and then factorizes the Hamiltonian $H=KhK^\dagger$ using the $KHK$ theorem, with all elements contained in $h$ being commutative~\cite{khaneja2001cartan}. As such, the time-evolution operator is reconstructed as
\begin{equation}
    e^{-iHt} = K e^{-iht} K^\dagger.
\end{equation}
The quantum circuit depth required to implement $e^{-iht}$ is independent of the total simulation time. QDSs based on Cartan decomposition are a good choice for a time-independent Hamiltonian. However, for a time-dependent Hamiltonian, Cartan decomposition should, in principle, be performed at each time step. This implies that the circuit depth increases with time evolution.

In this work, we extend Hamiltonian simulation via Cartan decomposition to time-dependent cases, with a focus on applying quantum computing to study the response of quantum systems to instantaneous external fields inducing spin diffusion and absorption spectra. This problem can be decomposed into a response part (where the external field is turned on) and a relaxed part (when the external field vanishes). The response part is handled with variational Hamiltonian simulation, in which the time-dependent wave function is approximated by applying parameterized quantum circuits onto an initial state. On the other hand, it should be pointed out that the variational part can be replaced by other methods, such as the adiabatic evolution method\cite{mc2024towards}. When the extenal field is a $\delta$-kick field, the state of system is often excited after this external perturbation. Subsequently, the relaxation part is evolved using the fast-forward propagation algorithm based on Cartan decomposition, which is a Lie algebra decomposition technique used to obtain the maximum Abelian set (MAS)~\cite{cirstoiu2020variational}. At each time step, the relevant physical quantities are measured to estimate the correlation function or spectra. Note that this new approach can be efficiently integrated with other techniques, such as stabilizer codes~\cite{bravyi2019simulation}, quantum state tomography~\cite{hai2023universal}, and more, to facilitate implementation on quantum computers. In this work, we denote this scheme as Variational-Cartan Quantum dynamics simulations (VCQDS).

\section{METHODOLOGY}
The Hamiltonian of the system including a time-dependent external field can be generally written as:
\begin{equation}\label{eq:time-dependent-ham}
    H(t) = H_0 + V(t),
\end{equation}
where $H_0$ is time-independent Hamiltonian and $V(t)$ is refered to an external field. In this work, we often consider a $\delta$-kick type external field, such external field is time localized and can be written as a $\delta$ function:
\begin{equation}\label{eq:kick-dependent-ham}
    V(t) =  E_0 \delta(t).
\end{equation}
However, the $\delta$ function in the time domain is not well-defined, so we approximate it with a Lorentzian function:
\begin{equation}\label{eq:external-field}
    \hat{V}(t) = \frac{E_0}{\pi}\frac{\Gamma}{\Gamma^2+t^2} \hat{D},
\end{equation}
where $\hat{D}$ is the interaction term and $E_0$ is the strength of the external field. Unless otherwise specified, the parameter $\Gamma$ in the Lorentz function is set to $0.25$ in this work. The coupling time $t_f$ between the external field and the system is controlled by $\Gamma$ and is generally much smaller than the total simulation time. Therefore, beyond the time domain $[0,t_f]$, the time evolution operator is considered  time-independent since the external field disappears. 

In this work, the $N$-body Hamiltonian $H$, after an appropriate mapping (e.g. Jordan-Wigner~\cite{jordan1928ber}
or Bravyi-Kitaev transformation~\cite{bravyi2002fermionic}), can be generally represented as:
\begin{equation}
    H(t) = \sum_i \lambda_i(t) P_i,
\end{equation} 
where $P_i$ is a tensor product in the form of $\{I,X,Y,Z\}^{\otimes N}$ and $\lambda_i(t)$ is the corresponding coefficient. Analogous to the variational quantum eigensolver ~\cite{PerMcCSha14,McCRomBab16}, the accuracy of VQDS depends on the expressivity of the circuit ansatz. A variety of parameterized quantum circuits, such as unitary coupled cluster~\cite{PerMcCSha14,SheZhaZha17,RomBabMcC18,LeeHugHea19} and hardware efficient ansatz~\cite{KanMezTem17,BarGonSok18,ZenFanLiu23} circuits, can be employed to represent the many-body wave function. In this work, we employ the Hamiltonian ansatz~\cite{wiersema2020exploring} to approximate the time-dependent wavefunction in the coupling process with the external field. After decomposing a given Hamiltonian into $M$ terms $H(t) = \sum_j^M \lambda_j(t) P_j$, the Hamiltonian ansatz of one layer is formulated as a product of a sequence of unitary exponentialized Pauli string operators:
\begin{equation}\label{eq:HA}
    U(\vec{\theta}) = \prod_{j}^M U_j\left(\theta_j\right) = 
    e^{-i \theta_M P_M} \cdots e^{-i \theta_1 P_1},
\end{equation}
where $\{\theta_j\}$ are real coefficients. In general, the accuracy of the Hamiltonian ansatz can be systematically improved by increasing the number of layers in parameterized quantum circuits:
\begin{equation}
    U(\vec{\theta}) = \prod_{L}^{N_L} U(\vec{\theta}_L).
\end{equation}
As mentioned above, the number of simulation steps in this process is minimal. Therefore, the issues of reference state selection and the ordering of terms in the Hamiltonian ansatz are effectively mitigated.

\begin{figure}[!htb]
    \centering
    \includegraphics[width=0.4\textwidth]{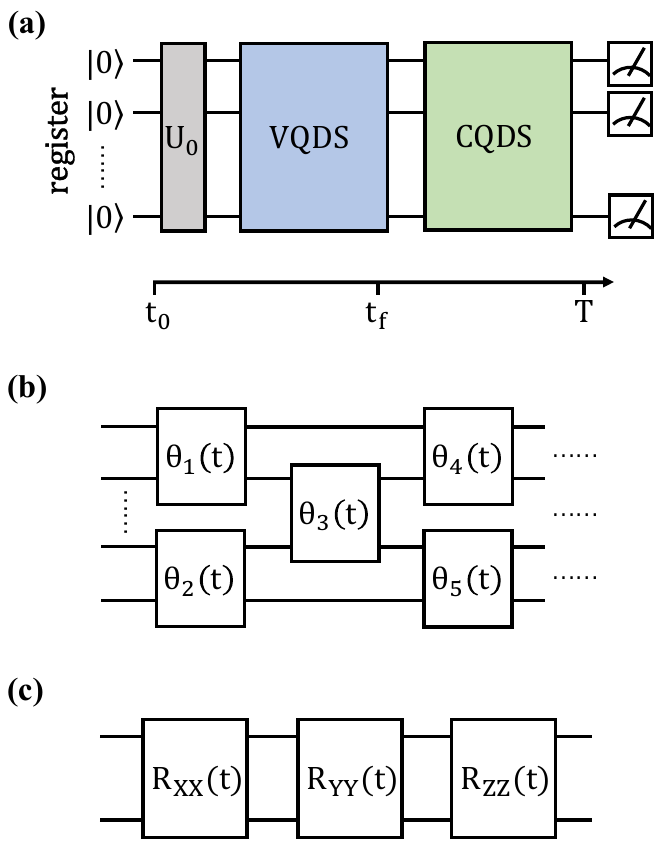} 
    \caption{(a) Flowchart of hybrid Hamiltonian simulaiton algorithm for quantum dynamics simulation (QDS) of photoexcitation processes, including initial state preparation, variational QDS (VQDS) and Cartan decomposition-based QDS; (b) A schematic quantum circuit for variational Hamiltonian simulation; (c) A schematic quantum circuit for Cartan decomposition-based Hamiltonian simulation.}
    \label{fig:flowchart}
\end{figure}

The entire dynamics process can be divided into two parts: the first part involves the kicked external field, where the system evolves under the external field. The variational Hamiltonian simulation algorithm can be used to simulate this time-dependent process. The second part involves the time evolution of the system under a time-independent Hamiltonian, which is treated using the Cartan decomposition-based Hamiltonian simulation algorithm. The flow chart of this hybrid Hamiltonian simulation is illustrated in Fig.~\ref{fig:flowchart}. 

\subsection{Variational Hamiltonian Simulation}
We start from a parameterized state:
\begin{equation}
    \ket{\psi(\vec{\theta}(t))} = U(\vec{\theta}(t)) \ket{\Psi_0},
\end{equation}
the variational principle can be introduced to simulate real time dynamics governed by the TDSE. Here, $\ket{\Psi_0}$ is an initial state and the trial state $\ket{\psi(\vec{\theta}(t))}$ that approximates the exact state $\ket{\Psi(t)}$ at time $t$ is prepared by applying a parameterized quantum circuit $U(\vec{\theta}(t))$ to the initial state. In case of ground state problems, the circuit parameters are optimized by minimizing the total energy~\cite{McCRomBab16,TilCheCao22}: 
\begin{equation}
    E=\braket{\psi(\vec{\theta}) |H| \psi(\vec{\theta})}.
\end{equation} 

For variational Hamiltonian simulation, the circuit parameters are optimized by minimizing the distance between the right and left hand sides of the TDSE (the McLachlan’s variational principle)~\cite{mclachlan1964variational} by
\begin{equation}
    \delta \left\lVert (\frac{\partial}{\partial t}+iH)
    \ket{\psi(\vec{\theta}(t))} \right
    \rVert = 0.
\end{equation}
Here, $\lVert |\Psi \rangle \rVert = \sqrt{\langle \Psi |\Psi \rangle}$ is the norm of quantum state $|\Psi\rangle$. Assuming $\theta$ to be complex, the evolution of variational parameters is evaluated by solving the linear equation
\begin{equation}
   \sum_{j} A_{i,j}\dot{\theta}_j = -i C_j. 
\end{equation}
When $\theta$ is real, the equation becomes
\begin{equation}
    \sum_{j} Re(A)_{i,j}\dot{\theta}_j = Im(C)_j.
\end{equation}
Here, the matrix elements of $\mathrm{A}$ and $\mathrm{C}$ are defined as
\begin{equation}       
\begin{split}
   A_{i,j} &= \frac{\partial\bra{\psi(\vec{\theta}(t))}}{\partial\theta_{i}}\frac{\partial\ket{\psi(\vec{\theta}(t))}}{\partial\theta_{j}},   \\
   C_j &= \frac{\partial\bra{\psi(\vec{\theta}(t))}}{\partial\theta_{j}} H \ket{\psi(\vec{\theta}(t))},
\end{split}
\end{equation}
$Re$ and $Im$ indicate real and imaginary parts, respectively. Note that the matrix $\mathrm{A}$ and vector $\mathrm{C}$ can be measured on quantum computer by using Hadamard test\cite{cleve1998quantum}.

In dynamics simulations, $|\psi(\vec{\theta}(t))\rangle$ differs from the targe state $|\Psi(t)\rangle$ by a global phase. In order to address this problem, one can represent $|\Psi(t)\rangle = e^{i\theta_0 t} |\psi(\vec{\theta}(t))\rangle$, resulting in a modified evolution equation~\cite{yuan2019theory}
\begin{equation}
   \sum_{j} M_{i,j}\dot{\theta}_j = V_j. 
\end{equation}
with
\begin{equation}
    \begin{split}
    M_{ij} &= Re(A)_{ij} + \frac{\partial\bra{\psi(\vec{\theta})}}{\partial\theta_{i}}\ket{\psi(\vec{\theta})} \frac{\partial\bra{\psi(\vec{\theta})}}{\partial\theta_{j}}\ket{\psi(\vec{\theta})}, \\
    V_i &= Im(C)_i + \bra{\psi(\vec{\theta})} \frac{\partial\ket{\psi(\vec{\theta})}} {\partial\theta_{j}}. 
    \end{split}
\end{equation}

The error of variational approaches to quantum dynamics can be measured by 
\begin{equation}
    \begin{split}
    &\left\lVert (\frac{\partial}{\partial t}+iH)
    \ket{\psi(\vec{\theta}(t))} \right
    \rVert^2 = \sum_{ij} Re(A)_{ij} \dot{\theta}_i \dot{\theta}_j \\
    &\quad\quad- 2 \sum_i Im(C)_i \dot{\theta}_i + \bra{\psi(\vec{\theta}(t))} H^2 \ket{\psi(\vec{\theta}(t))}.
    \end{split}
\end{equation}
The accuracy of the simulation is thereby assessed by this distance between the optimized state and the true state. 

\subsection{Cartan Decomposition-based Hamiltonian Simulation}\label{sec:Cartan}
Hamiltonian simulation via Cartan decomposition is appealing for long time evolution of the quantum state under a time-independent Hamiltonian with fixed circuit depth. Cartan decomposition is a useful technique for decomposing of a unitary operator into a sequence of 1- and 2-qubit operations~\cite{EarPac05}. In this work, Cartan decomposition is used to factorize the time evolution operator under a time-independent Hamiltonian $H_0$. The closure (under commutation and linear combination) of the set of Pauli terms $\{P_i\}$ in $H_0$ becomes a Lie subalgebra $\mathfrak{g} \subset \mathfrak{s u}\left(2^n\right)$, which has a compact semi-simple Lie subgroup $\boldsymbol{G} \subset S U\left(2^n\right)$ in exponential map. A Cartan decomposition on $\mathfrak{g}$ is defined as finding an orthogonal split $\mathfrak{g}=\mathfrak{k} \oplus \mathfrak{m}$, satisfying
\begin{equation}
[\mathfrak{k}, \mathfrak{k}] \subset \mathfrak{k}, \quad[\mathfrak{m}, \mathfrak{m}] \subset \mathfrak{k}, \quad[\mathfrak{k}, \mathfrak{m}]=\mathfrak{m}.   
\end{equation}
This decompostion is labelled as $(\mathfrak{m}, \mathfrak{k})$. The Cartan subalgebra of this decomposition is defined as one of the maximal Abelian subalgebras of $\mathfrak{m}$, named $\mathfrak{h}$. 

In practice, the Lie subalgebra is partitioned by an involution. A Cartan involution is a Lie algebra homomorphism  $\theta: \mathfrak{g} \rightarrow \mathfrak{g}$, which satisfies $\theta(\theta(g))=g$ for any $g \in \mathfrak{g}$. This homomrophism preserves all commutators and distinguishes $\mathfrak{k}$ and $\mathfrak{m}$ by $\theta(\mathfrak{k})=\mathfrak{k}$ and $\theta(\mathfrak{m})=-\mathfrak{m}$.
Given a Cartan decomposition $\mathfrak{g}=\mathfrak{k} \oplus \mathfrak{m}$, for any element $m \in \mathfrak{m}$ there exists a $K \in e^{\mathfrak{k}}$ and an $h \in \mathfrak{h}$, such that
\begin{equation}
 m=K h K^{\dagger}.   
\end{equation}

Note that $H_0\in\mathfrak{m}$ so that one needs to find an appropriate $K$ that gives $H_0=K h K^\dagger$. In the Lie algebra, one can define a symmetric bilinear form as Killing form. For $\mathfrak{s u}\left(2^n\right)$, it is able to define the Killing form as (given $X, Y \in \mathfrak {g}$ viewed in their fundamental matrix representation):
\begin{equation}
    \kappa(X,Y) = \left\langle X, 
    Y\right\rangle = 2^{n+1} \mathrm{Tr}(XY).
\end{equation}
we only choose the trace part of standard Killing form in our work. The $K$ is determined by finding a local minimum of
\begin{equation}\label{eq:fk}
    f(K)=\left\langle K v K^{\dagger}, 
    H_0\right\rangle,
\end{equation}
where $\braket{.,.}$ denotes the Killing form and $v \in \mathfrak{h}$ is an element whose exponential map is dense in $e^{i \mathfrak{h}} $. For example, given $v=\sum_i \gamma^i h_i$, $h_i$ forms a basis for $\mathfrak{h}$, and $\gamma^i$ is an arbitrary number. At a local minimum of $f(K)$, $K_0^{\dagger} H_0 K_0=h \in \mathfrak{h}$ holds, which determines $h$ and thus completes the decomposition.
The form of $K$ can be chosen as~\cite{KokSteWan22}:
\begin{equation}
K= e^{\sum_i i a_i k_i} \quad \mathrm{or} \quad K=\prod_i e^{i a_i k_i},
\end{equation}
where $k_i$ is a Pauli string basis for $\mathfrak{k}$. Note that these two forms are equivalent due to compact Lie group and $\mathfrak{k}$ is closed under commutation. It is clear that finding $K$ is the computational bottleneck in the Cartan decomposition. However, as shown in Fig.~\ref{fig:flowchart}, the quantum circuit of $K$ is time-independent in the sense that its parameters are only required to determine once. Each Pauli string in $h$ commutes since $h$ is an element of an Abelian group. As such, the first-order Trotter expansion of $e^{-iht}$ is exact, implying that the circuit depth will not increase as time evolution. 

\section{RESULTS}
In this work, we apply the VCQDS approach to study quantum dynamics of model and molecular systems after photoexcitation. Molecular ground-state properties are computed with PySCF~\cite{ns2018booth} and the VQE calculations are performed using Q$^2$Chemistry~\cite{q2chemistry}. The Cartan decomposition is carried out according to the method documented in Ref.~\citenum{KokSteWan22}. In VCQDS, we employ Hamiltonian ansatz consisting of multiple layers of the first-order Trotterization decomposition of the time-evolution operator (Eq.~\eqref{eq:HA}) to represent the time-dependent wave function, unless otherwise specified.

\begin{figure*}[!htb]
    \centering
    \includegraphics[width=0.8\linewidth]{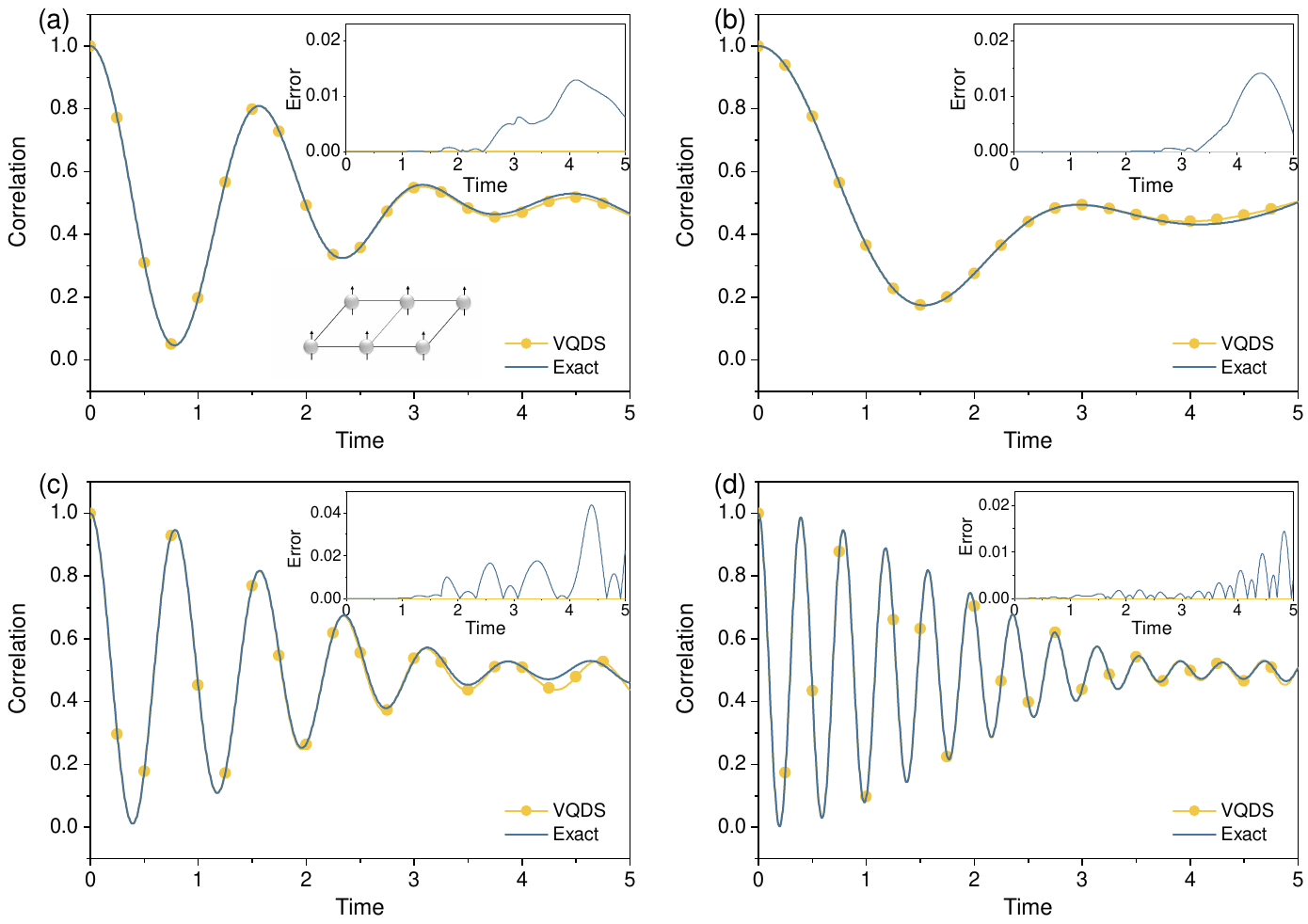} 
    \caption{Numerical simulations of the 2D Ising Model using the variational approach, named VQS. We consider time evolution from $t = 0$ to $5$ with $\delta t = 0.005$, and measure the nearest-neighbor correlations $C = \frac{1}{7}\sum_{<i,j>} Z_i Z_j$. We set the corresponding parameters to (a) $J/d = 1$ (b)$ J/d = 2$ (c) $J/d = 0.5$ (d) $J/d = 0.25$. The insets shows the errors of the correlation function with respect to the exact results.}
    \label{fig:Ising_model}
\end{figure*}

\subsection{Variational Hamiltonian Simulation}
Firstly, we numerically test the variational approach for dynamics simulations of a 6-qubit 2D Ising model in a transverse field. The Hamiltonian is written as
\begin{equation}
  H = \frac{J}{4}\sum_{<i,j>} Z_iZ_j + d\sum_i^{N_{s}} X_i,
\end{equation}
where the Pauli operators $X_i$, $Y_i$, and $Z_i$ act on the $i$-th site and $\braket{i,j}$ represents the nearest-neighbor pair of sites. The geometric structure is shown in Fig.~\ref{fig:Ising_model} (a). $N_s$ is the total number of sites, that is $N_s=6$. The initial state is prepared in a product state $\ket{\psi} = \ket{0}^{\otimes 6}$, and the Hamiltonian ansatz is defined as 
\begin{equation}
U = U_h(\vt_1) U_h(\vt_2) U_h(\vt_3) U_s(\vt_4) U_h(\vt_5) U_h(\vt_6) U_h(\vt_7) U_s(\vt_8)
\end{equation}
with  
\begin{equation}
\begin{split}
U_h (\vt_k) = &e^{-i \theta_{k,1}(Z_5Z_6)} e^{-i \theta_{k,2}(Z_3Z_5+Z_4Z_6)} e^{-i \theta_{k,3}(Z_3Z_4)} \\
&e^{-i \theta_{k,4}(Z_1Z_3+Z_2Z_4)} e^{-i \theta_{k,5}(Z_1Z_2)}
\end{split}
\end{equation}
and 
\begin{equation}
    U_s (\vt_k) = \prod_{l=1}^6 e^{-i \theta_{k,l} X_{l}}
\end{equation}
as introduced in Ref.~\citenum{SunEndLin21}.

The time evolution of the correlation function
\begin{equation}
     C = \frac{1}{7} \sum_{<i,j>} Z_i Z_j
\end{equation}
for different coupling strengths, including $J/d$=1, 2, 0.5 and 0.25, are shown in Fig.~\ref{fig:Ising_model}. We compare the dynamical nearest-neighbor correlations computed using the variational approach with respect to the exact results. It is clear that numerical errors of VQDSs enlarge as the simulation time increases due to both the numerical implementation error $\epsilon_I$ and algorithmic error $\epsilon_A$ ~\cite{SunEndLin21}. In cases of $J/d$=1, 2, and 0.25, the maximal errors of the correlation functions computed using the variational approach are $\sim$0.01, that is the maximal relative errors are $\sim$ 2\% during time evolution from $t$=0 to 5. In case of $J/d$=0.5, the maximal relative error is as large as 0.04. 

On a NISQ quantum computer, the implementation error mainly comes from the imprecise estimation of $M$ and $V$ owing to the presence of both physical and shot noise. The algorithmic error mainly comes from a finite time step, a commonly existing problem in dynamics simulation, and an approximate variational ansatz to represent the exact quantum state. In fact, the error bounds related to the statistical average and time step are $\sim\frac{T}{\sqrt{N_\mathrm{M}}}$ and $\sim\sqrt{\delta t} T$, respectively, where $N_M$ is the number of measurements. To achieve the required accuracy, one can choose $N_\mathrm{M}$ as $\sim \frac{T^2}{\epsilon_{I}^2}$ and $\delta t$ as $\sim \frac{\epsilon_{A}^2}{T^2}$, which is consistent with Chebyshev’s inequality. At the same time, the number of time steps $N_\mathrm{T} = \frac{T}{\delta t}\sim T^3$ increases rapidly as the total simulation time $T$, resulting in an unaffordable computational cost in long-time dynamics simulations. The accuracy of the ansatz can be improved using the adaptive variational scheme~\cite{barison2021efficient,ZhaSunYua23} while this scheme may lead to increasing circuit depth as the time evolution. In the case of different $J/d$ parameters, the variational approach is sufficiently accurate to perform short-time dynamics. 

\subsection{Molecular Absorption Spectra}
Designing organic electronic devices have attracted broad interest over past decades\cite{anthony2006functionalized}. Acenes are an important ingredient in two classes of electronic devices: field-effect transistors (FETs, also knownas thin-film transistors, TFTs) and organic light-emitting diodes (OLEDs). The colour of acenes and Heteroacenes in the visible spectrum is primarily determined by the energy gap between their highest occupied molecular orbital (HOMO) and lowest unoccupied molecular orbital (LUMO). In principle, the active space that consists of the HOMO and LUMO is able to qualitatively describe the lowest electronic excitation of polyacenes. While, in order to quantatively describe their spectra, we need to include the dynamic correlation effect beyond the active space. As such, all the computed spectra of polyacenes have been shifted by -1.9 eV, which accounts for the difference between the complete active space (CAS) configuration interaction (CI) and CAS second-order perturbation theory~\cite{HuaCaiLi22}.

\begin{figure}[!htb]
    \centering
    \includegraphics[width=0.9\linewidth]{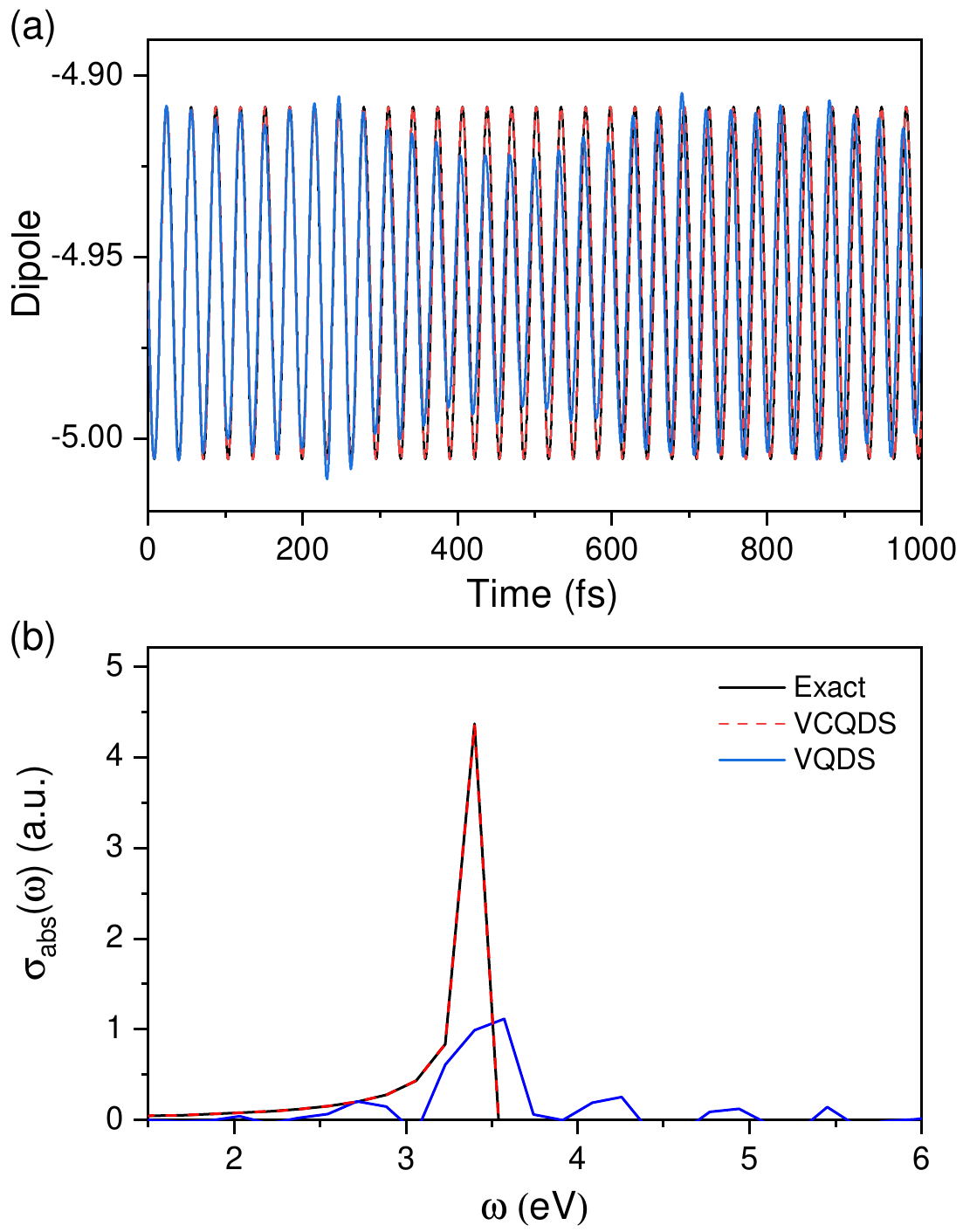} 
    \caption{(a) Time evolution of the dipole moment for anthracene computed using the exact time-evoluation, VCQDS and VQDS approaches. (b) Absorption spectra simulated by the Fourier transformation of the dipole moments. The parameter $\gamma$ of the full width at half maximum is 0.001.}
    \label{fig:polyacenes}
\end{figure}

We apply the VCQDS algorithm to simulate molecular absorption spectra for a series of polyacenes, including naphthalene, anthracene, tetracene, and pentacene, which are polycyclic aromatic hydrocarbons composed of linearly fused benzene rings. The absorption spectra is an important tool to characterize the optical property of polyacenes. As the number of fused benzene rings increases, the first absorption peak exhibits an an obvious color change from blue to green to yellow due to the evolution of electronic structure~\cite{clar1964polycyclic}. 

Consider a CAS(2o,2e) model, the following simulations involves only four molecular spin orbitals. The one- and two-electron integrals for constructing the second-quantized Hamiltonian were extracted from Hartree-Fock calculations. All calculations were performed with the 6-31G* basis.
The $\delta$-kick field strength $E_0$ is set to $0.01$ and the coupling operator is
\begin{equation}
    \hat{D} = \hat{\mu},
\end{equation}
with $\hat{\mu}$ being the diple moment of molecules.

The absorption spectra can be obtained by calculating the dynamical polarizability tensor from the time evolution of the dipole moments. The elements of the dynamical polarizability tensor can be given by
\begin{equation}
\alpha_{v v^{\prime}}=\frac{d_v(\omega)}{E_{v^{\prime}}(\omega)},
\end{equation}
where $v$ and $v^\prime \in \{x, y, z\}$ denote different directions, and $d_v(\omega)$ and $E_v(\omega)$ denote the Fourier transformations of the dipole moment and applied electric fields in the $v$ direction, respectively. The optical absorption cross-section is
\begin{equation}
\sigma_{abs}(\omega)=\frac{4 \pi \omega}{c} \operatorname{Im}[\alpha(\omega)].
\end{equation}

Figure.~\ref{fig:polyacenes} (a) shows the  evolution of the dipole moment for anthracene over time simulated using the exact time-evoluation, VCQDS and VQDS approaches. The VCQDS approach is able to reproduce the exact results over 1000 fs. The dipole moment from the VCQDS simulation exhibits strict periodicity, and its intensity is almost constant throughout the simulation process, indicating that the simulation process has high fidelity. It is evident that the VQDS approach is also able to recover the exact results at a short time scale while for a long time evoluation, the VQDS approach exhibits significant deviation of the dipole moment with respect to the exact approach. The Fourier transformation of the dipole moments is illustrated in Figure.~\ref{fig:polyacenes} (b). The absorption spectra from the VCQDS simulation yields a unique peak that originates from the electron excitation from the HOMO to LUMO. In contrast, the VQDS simulation produces many small meaningless peaks, which results from the inaccurate oscillation in the dipole moment. The position of the highest peak also deviates from the accurate result.

\begin{figure}
    \centering
    \includegraphics[width=0.9\linewidth]{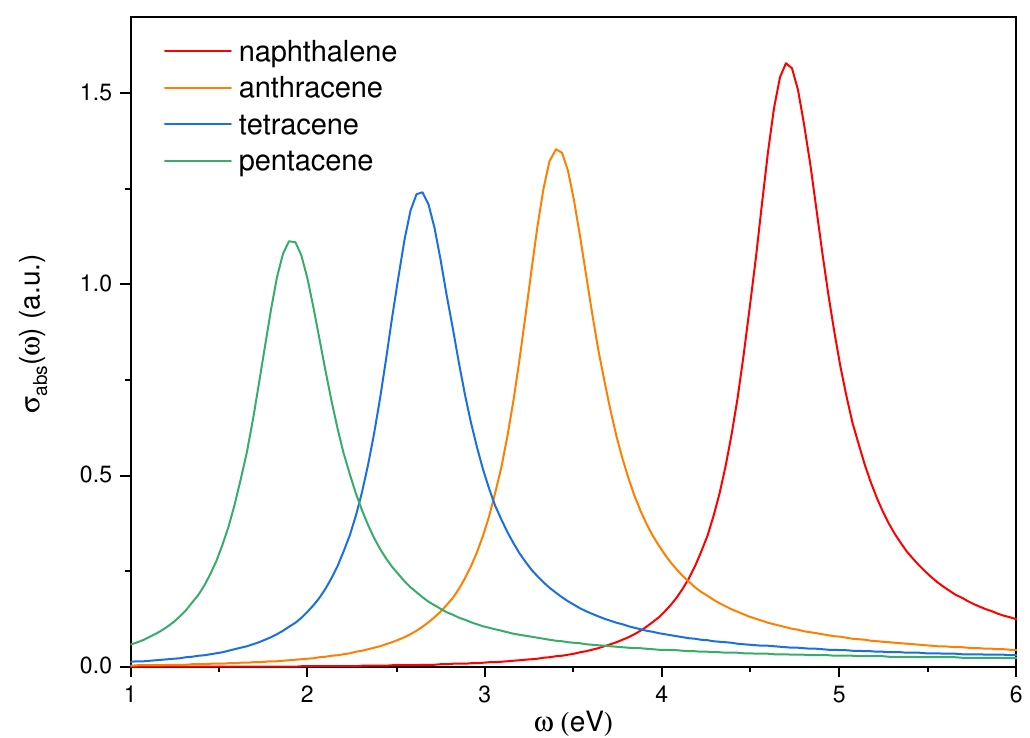} 
    \caption{Absorption spectra simulated using the VCQD approach for naphthalene, anthracene, tetracene, and pentacene. The parameter $\gamma$ of the full width at half maximum is 0.01. All the computed spectra have been shifted by -1.9 eV estimated for dynamic electron correlation.}
    \label{fig:polyacene_spectra}
\end{figure}

Figure.~\ref{fig:polyacene_spectra} shows the absorption spectra $\sigma_{\mathrm{abs}}(\omega)$ simulated using the VCQD approach for naphthalene, anthracene, tetracene, and pentacene. As the number of benzene rings increases, the wavelength of the signal gradually increases, which is consistent with the trend of redshift observed in the  experimental spectral data~\cite{clar1964polycyclic}. In addition, the intensity of the peaks gradually decreases from anphthalene to pentacene.

\subsection{Spin evolution of the Heisenberg model}
We apply the hybrid Hamiltonian simulation approach to study one-dimensional Heisenberg models with competing interactions between nearest neighbor sites in periodic chains. The elementary excitation in the Heisenberg model is known as a magnon, which behaves as a boson after the Holstein-Primakoff (H-P) transformation. Simulating bosonic excitations is challenging using the diagonalization approach because there is no upper limit on the number of bosons. Alternatively, one can compute the dynamical correlation function to extract the spectra. The spin system is a natural candidate for simulation on a quantum computer~\cite{pedernales2014efficient,FraFreKem20}. Here, we require to compute spin susceptibilities, which can be defined as:
\begin{equation}
    \braket{S_z^i(t)} - \braket{S_z^i(0)}\equiv \delta S_z^i(t) = \sum_j\int_0^{+\infty}  \chi_{ij}(t-t^\prime)V_j(t^\prime)dt^\prime.
\end{equation}
When an external field acts locally on a single qubit, the elements of the dynamical correlation function tensor can be given by
\begin{equation}\label{eq:chi}
    \chi_{ij}(\omega) = \frac{\delta S_z^i(\omega)}{V_j(\omega)} = \frac{\delta S_z^i(\omega)}{E_0},
\end{equation}
where $S_z^i(\omega)$ and $V_j(\omega)$ denote the Fourier transformations of the z-direction spin component on the $i^{th}$ qubit and applied electric fields in the z-direction on the $j^{th}$ qubit, respectively. In linear response theory, the susceptibility is closely related to a retarded Green function, so the correspoding relation for the two-point correlation functions ($C_{ij}(\omega)$) is~\cite{fetter2012quantum}:
\begin{equation}\label{eq:retard-time-order}
\begin{split}
   &Re\chi_{ij}(\omega) = Re C_{ij}(\omega),   \\
   &Im\chi_{ij}(\omega)sgn\omega = Im C_{ij}(\omega).
\end{split}
\end{equation}

\begin{figure}[htb]
    \centering
    \includegraphics[width=0.9\linewidth]{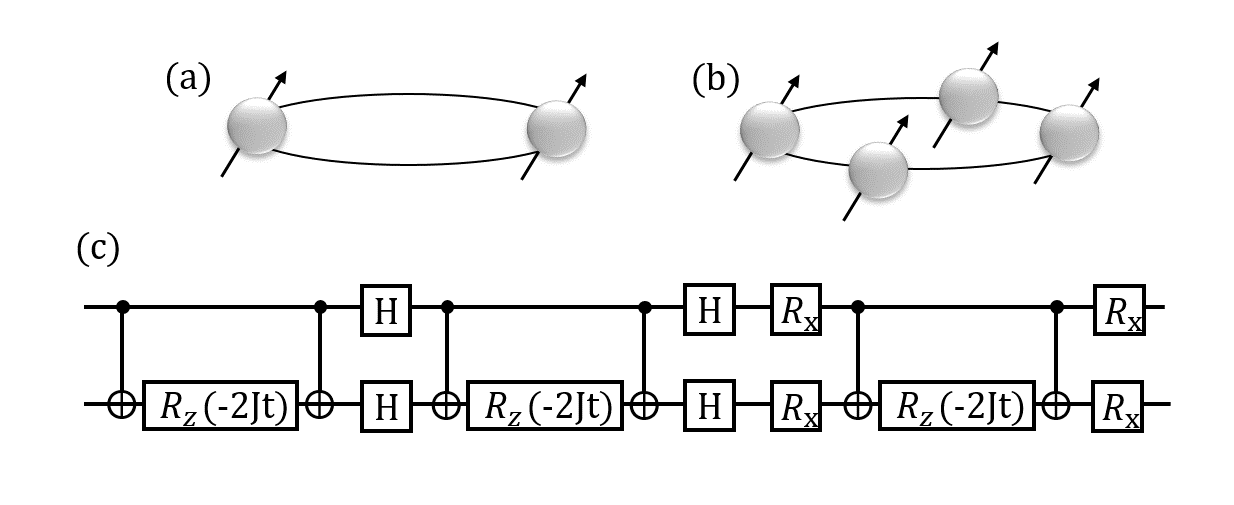} 
    \caption{Geometric structures of the Heisenberg model: (a) two-site (b) four-site. (c) Quantum circuit for implementing $\exp(-i H t)$. $H$ is the bare two-site Heisenberg model Hamiltonian. $"X"$ denotes Pauli "$X$", and $R_{\alpha}(\theta)$ denotes a rotation by $\theta$ on the $\alpha$-axis.}
    \label{fig:Hescircuit}
\end{figure}

\begin{figure}[htb]
    \centering
    \includegraphics[width=1\linewidth]{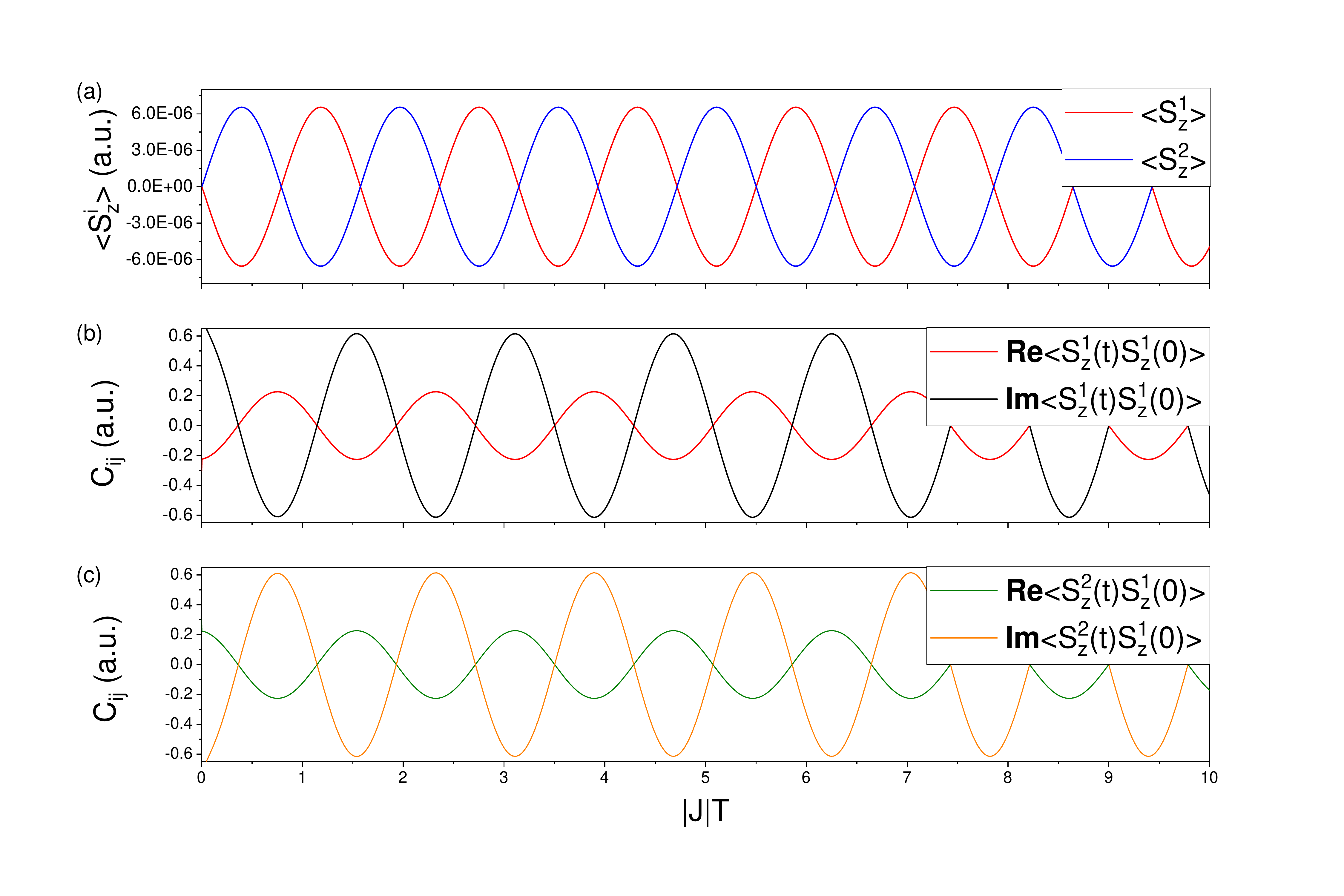} 
    \caption{(a) Time evolution of spins in the Z direction for the two-site antiferromagnetic Heisenberg model. Real and imaginary parts of (b) $\braket{S^1_z(t)S^1_z(0)}$ and (c) $\braket{S^2_z(t)S^1_z(0)}$ for the two-site antiferromagnetic Heisenberg model. $|J|$ is the coupling strength, $T$ is the simulation time.}
    \label{fig:Heisenberg2site}
\end{figure}

Here, the electron spin is set to one-half, so the spin operator in the spin basis becomes the Pauli matrix. The initial Hamiltonian reads
\begin{equation}
    H= \sum_{\braket{i,j}}J_{x}X_iX_j+J_{y}Y_iY_j+J_{z}Z_iZ_j.
\end{equation}
First, we consider the antiferromagnetic Heisenberg model on two sites with all the coupling coefficients are positive and equal to $|J|$. The geometric structure of the two-sites periodic Heisenberg model is shown in Fig.~\ref{fig:Hescircuit}(a). This Hamiltonian is a special example, in which all of terms are commutative with each other. As such, the time envolution operator is expanded as
\begin{equation}
    e^{-i H t} = e^{-iJ_xtX_1X_2}e^{-iJ_ytY_1Y_2}e^{-iJ_ztZ_1Z_2},
\end{equation}
every exponential term in the time envolution operator directly corresponds to the 'H(X)-CNOT-R-CNOT-H(X)' circuit structure and the whole quantum circuit is shown in Fig.~\ref{fig:Hescircuit}(c). The initial state is chosen to be
\begin{equation}
    \ket{\Psi_0} = \frac{1}{\sqrt{2}}(\ket{\uparrow\downarrow} - \ket{\downarrow\uparrow})
\end{equation}
which differs the ground state of the antiferromagnetic Heisenberg model by an arbitrary global phase.
We apply a $\delta$-kick external field on the first site, so the $\hat{D}$ is $Z_1$ and set field strength $E_0$ to $10^{-5}$. The time evolution of z-component of the spin on each site in the two-site Heisenberg model is shown in Fig.~\ref{fig:Heisenberg2site} (a). It is evident that the spin z-component exhibits a strict periodic oscillatory behavior between two sites and this oscillation satisfies convervation of spin z-component. Following Eq.~\ref{eq:chi} and Eq.~\ref{eq:retard-time-order}, we can obtain the corresponding two-point correlation function between different sites, as shown in Fig.~\ref{fig:Heisenberg2site} (b) and (c), respectively. One can see that there is an evident symmetry and phase difference relationship between the two correlation functions. The phase difference between them accurately reflects the wave vector and the intensity of the spin wave, which can be obtained through Fourier transform to yield the corresponding magnon spectrum~\cite{pedernales2014efficient}.

We also apply the hybrid Hamiltonian simulation approach to study the four-site antiferromagnetic periodic Heisenberg model. The corresponding geometric structure of the four-site periodic Heisenberg model is shown in Fig.~\ref{fig:Hescircuit} (b). The difference between two-site and four-site Heisenberg models lies in the non-commutativity between two certain terms in the four-site Heisenberg Hamiltonian. As discussed in Sec.~\ref{sec:Cartan}, the Cartan decomposition can separate out the mutually commuting parts, which makes the time in the time evolution operator become a parameter in the circuit, and thus one can carry out the time evolution without Trotterization. 

\begin{figure}[htb]
    \centering
    \includegraphics[width=0.9\linewidth]{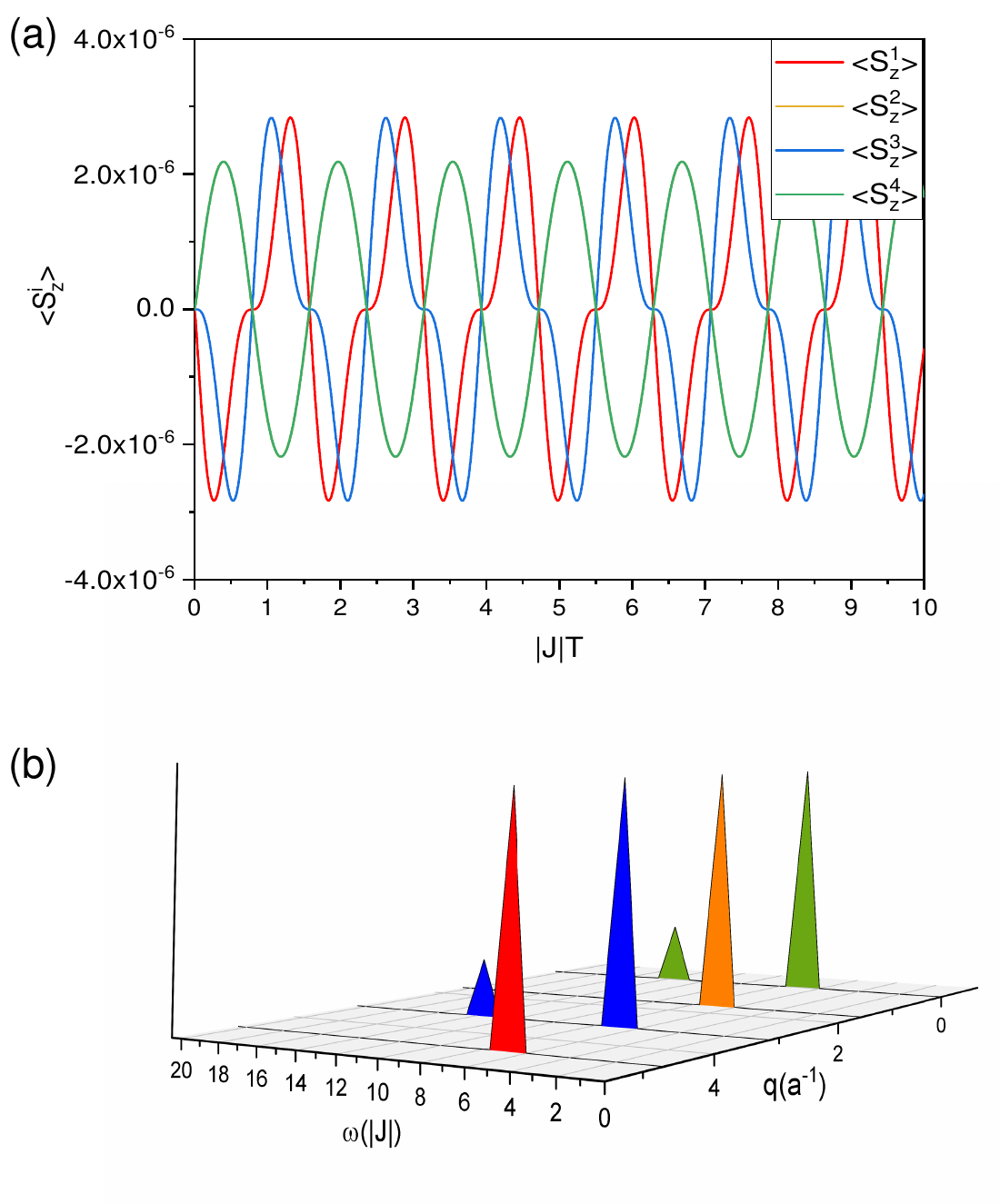} 
    \caption{(a) Time evolution of spins in the Z direction for four-site antiferromagnetic Heisenberg model. (b) Magnon spectra of four-site antiferromagnetic Heisenberg model response to a $\delta$-kick external field. $|J|$ is the coupling strength, $q$ is the reciprocal lattice vector.}
    \label{fig:Heisenberg4site}
\end{figure}

We choose the ground state of the four site antiferromagnetic Heisenberg model $\ket{\Psi_0}$ as the 
initial state:
\begin{equation}
    \begin{split}
    \ket{\Psi_0} &= \frac{1}{\sqrt{12}}\sum(\ket{\downarrow\uparrow\downarrow\uparrow}+2\ket{\downarrow\downarrow\uparrow\uparrow}),
    \end{split}
\end{equation}
where the summation symbol represents summing over all possible configurations of the corresponding two-spin distributions in the periodic chain. Similarly, the $\delta$-kick external field acts on the first qubit. Fig.~\ref{fig:Heisenberg4site} (a) shows the time evolution of z-component of the spin on each site for four-site Heisenberg model. We can see the z-component of the spin at the second and fourth sites coincide in the time evolution, which conforms to the periodic boundary conditions. This is also reflected in the corresponding ZZ spin correlation functions, where the correlation functions $\braket{S_z^2(t)S_z^1(0)}$ and $\braket{S_z^4(t)S_z^1(0)}$ are exactly identical. Moreover, we can observe that there is significant phase difference between correlation functions of $\braket{S_z^1(t)S_z^1(0)}$ and $\braket{S_z^3(t)S_z^1(0)}$, analogous to that in the two-site Heisenberg model. The low-energy excitations of the periodic Heisenberg chain are magnons. A reconstruction of the magnon spectrum requires to perform a Fourier transform of the correlation functions from real space to momentum space and from time to frequency. The magnon spectra can be computed as:
\begin{equation}
    \left\langle S_x^i(t) S_x^j(0)\right\rangle=\sum_q A_q e^{i q\left(r_i-r_j\right)} e^{-i \omega_q t} .
\end{equation}
Fig.~\ref{fig:Heisenberg4site} shows the magnon spectra for the four-site Heisenberg model, which provides collective excitations and dynamic properties of spin systems. The peaks in the magnon spectra are the excitation energies from antiferromagnetic ground state to the first excited state.

\section{Conclusion}
In conclusion, we propose a hybrid Hamiltonian simulation algorithm --- VCQDS for long time dynamics simulations of systems response to a short time external field. Variational Hamiltonian simulation using fixed-depth quantum circuits is employed to handle the time-dependent Hamiltonian consisting of an external field. Cartan decomposition-based Hamiltonian simulation is able to exactly dealing with time evolution of the system after the external field disappers. Since the Cartan decomposition is exact, it strictly maintain the periodicity in the time-dependent evolution of physical quantities, ensuring the correctness of the computed spectra. Moreover, this method does not require Trotterization, enabling us to take arbitrarily small step sizes while maintaining the overall error without increasing with the number of steps. This hybrid Hamiltonian simulation algorithm is applied to study the photoexcitation processes of the Heisenberg model and polyacenes with quite acceptable accuracy.

While it is necessary to point out that Cartan decomposition does not have good scalability because we only utilize the anti-commutativity of the operators in the Hamiltonian, resulting a rapidly expanding Lie algebra space as the degrees of freedom in the system increases. One possible solution is to leverage the property of the initial wave function to pose certain restrictions on the Lie algebra space generated from the system Hamiltonian in the sense that the dimension of the Lie algebra space is significantly reduced. On the other hand, it is also possible to explore the localized structure of the Hamiltonian to reduce the compuational complexity. 

\section{Acknowledgments}
This work was supported by Innovation Program for Quantum Science and Technology (2021ZD0303306), the National Natural Science Foundation of China (22073086, 22288201), Anhui Initiative in Quantum Information Technologies (AHY090400), and the Fundamental Research Funds for the Central Universities (WK2060000018).

\section{Conflicts of interest}
There are no conflicts of interest to declare.

\footnotesize{
\bibliography{achemso}

\providecommand{\latin}[1]{#1}
\makeatletter
\providecommand{\doi}
  {\begingroup\let\do\@makeother\dospecials
  \catcode`\{=1 \catcode`\}=2 \doi@aux}
\providecommand{\doi@aux}[1]{\endgroup\texttt{#1}}
\makeatother
\providecommand*\mcitethebibliography{\thebibliography}
\csname @ifundefined\endcsname{endmcitethebibliography}  {\let\endmcitethebibliography\endthebibliography}{}
\begin{mcitethebibliography}{60}
\providecommand*\natexlab[1]{#1}
\providecommand*\mciteSetBstSublistMode[1]{}
\providecommand*\mciteSetBstMaxWidthForm[2]{}
\providecommand*\mciteBstWouldAddEndPuncttrue
  {\def\EndOfBibitem{\unskip.}}
\providecommand*\mciteBstWouldAddEndPunctfalse
  {\let\EndOfBibitem\relax}
\providecommand*\mciteSetBstMidEndSepPunct[3]{}
\providecommand*\mciteSetBstSublistLabelBeginEnd[3]{}
\providecommand*\EndOfBibitem{}
\mciteSetBstSublistMode{f}
\mciteSetBstMaxWidthForm{subitem}{(\alph{mcitesubitemcount})}
\mciteSetBstSublistLabelBeginEnd
  {\mcitemaxwidthsubitemform\space}
  {\relax}
  {\relax}

\bibitem[Nelson \latin{et~al.}(2020)Nelson, White, Bjorgaard, Sifain, Zhang, Nebgen, Fernandez-Alberti, Mozyrsky, Roitberg, and Tretiak]{NelWhiBjo20}
Nelson,~T.~R.; White,~A.~J.; Bjorgaard,~J.~A.; Sifain,~A.~E.; Zhang,~Y.; Nebgen,~B.; Fernandez-Alberti,~S.; Mozyrsky,~D.; Roitberg,~A.~E.; Tretiak,~S. Non-adiabatic excited-state molecular dynamics: Theory and applications for modeling photophysics in extended molecular materials. \emph{Chem.\ Rev.} \textbf{2020}, \emph{120}, 2215--2287\relax
\mciteBstWouldAddEndPuncttrue
\mciteSetBstMidEndSepPunct{\mcitedefaultmidpunct}
{\mcitedefaultendpunct}{\mcitedefaultseppunct}\relax
\EndOfBibitem
\bibitem[Miessen \latin{et~al.}(2023)Miessen, Ollitrault, Tacchino, and Tavernelli]{MieOllTac23}
Miessen,~A.; Ollitrault,~P.~J.; Tacchino,~F.; Tavernelli,~I. Quantum algorithms for quantum dynamics. \emph{Nat. Comput. Sci.} \textbf{2023}, \emph{3}, 25--37\relax
\mciteBstWouldAddEndPuncttrue
\mciteSetBstMidEndSepPunct{\mcitedefaultmidpunct}
{\mcitedefaultendpunct}{\mcitedefaultseppunct}\relax
\EndOfBibitem
\bibitem[Feynman(1982)]{feynman2018simulating}
Feynman,~R.~P. Simulating Physics with Computers. \emph{Int. J. Theor. phys} \textbf{1982}, \emph{21}\relax
\mciteBstWouldAddEndPuncttrue
\mciteSetBstMidEndSepPunct{\mcitedefaultmidpunct}
{\mcitedefaultendpunct}{\mcitedefaultseppunct}\relax
\EndOfBibitem
\bibitem[Lloyd(1996)]{lloyd1996universal}
Lloyd,~S. Universal quantum simulators. \emph{Science} \textbf{1996}, \emph{273}, 1073--1078\relax
\mciteBstWouldAddEndPuncttrue
\mciteSetBstMidEndSepPunct{\mcitedefaultmidpunct}
{\mcitedefaultendpunct}{\mcitedefaultseppunct}\relax
\EndOfBibitem
\bibitem[Wiesner(1996)]{wiesner1996simulations}
Wiesner,~S. Simulations of many-body quantum systems by a quantum computer. \emph{arXiv preprint quant-ph/9603028} \textbf{1996}, \relax
\mciteBstWouldAddEndPunctfalse
\mciteSetBstMidEndSepPunct{\mcitedefaultmidpunct}
{}{\mcitedefaultseppunct}\relax
\EndOfBibitem
\bibitem[Zalka(1998)]{zalka1998efficient}
Zalka,~C. Efficient simulation of quantum systems by quantum computers. \emph{Fortschr Phys.} \textbf{1998}, \emph{46}, 877--879\relax
\mciteBstWouldAddEndPuncttrue
\mciteSetBstMidEndSepPunct{\mcitedefaultmidpunct}
{\mcitedefaultendpunct}{\mcitedefaultseppunct}\relax
\EndOfBibitem
\bibitem[Tacchino \latin{et~al.}(2020)Tacchino, Chiesa, Carretta, and Gerace]{tacchino2020quantum}
Tacchino,~F.; Chiesa,~A.; Carretta,~S.; Gerace,~D. Quantum computers as universal quantum simulators: state-of-the-art and perspectives. \emph{Adv. Quantum Technol.} \textbf{2020}, \emph{3}, 1900052\relax
\mciteBstWouldAddEndPuncttrue
\mciteSetBstMidEndSepPunct{\mcitedefaultmidpunct}
{\mcitedefaultendpunct}{\mcitedefaultseppunct}\relax
\EndOfBibitem
\bibitem[Daley \latin{et~al.}(2022)Daley, Bloch, Kokail, Flannigan, Pearson, Troyer, and Zoller]{daley2022practical}
Daley,~A.~J.; Bloch,~I.; Kokail,~C.; Flannigan,~S.; Pearson,~N.; Troyer,~M.; Zoller,~P. Practical quantum advantage in quantum simulation. \emph{Nature} \textbf{2022}, \emph{607}, 667--676\relax
\mciteBstWouldAddEndPuncttrue
\mciteSetBstMidEndSepPunct{\mcitedefaultmidpunct}
{\mcitedefaultendpunct}{\mcitedefaultseppunct}\relax
\EndOfBibitem
\bibitem[Lee \latin{et~al.}(2022)Lee, Qin, Raussendorf, Sela, and Scarola]{PhysRevResearchMeasurementLee}
Lee,~W.-R.; Qin,~Z.; Raussendorf,~R.; Sela,~E.; Scarola,~V.~W. Measurement-based time evolution for quantum simulation of fermionic systems. \emph{Phys. Rev. Res.} \textbf{2022}, \emph{4}, L032013\relax
\mciteBstWouldAddEndPuncttrue
\mciteSetBstMidEndSepPunct{\mcitedefaultmidpunct}
{\mcitedefaultendpunct}{\mcitedefaultseppunct}\relax
\EndOfBibitem
\bibitem[Smith \latin{et~al.}(2019)Smith, Kim, Pollmann, and Knolle]{smith2019simulating}
Smith,~A.; Kim,~M.; Pollmann,~F.; Knolle,~J. Simulating quantum many-body dynamics on a current digital quantum computer. \emph{npj\ Quantum\ Inf.} \textbf{2019}, \emph{5}, 106\relax
\mciteBstWouldAddEndPuncttrue
\mciteSetBstMidEndSepPunct{\mcitedefaultmidpunct}
{\mcitedefaultendpunct}{\mcitedefaultseppunct}\relax
\EndOfBibitem
\bibitem[Georgescu \latin{et~al.}(2014)Georgescu, Ashhab, and Nori]{georgescu2014quantum}
Georgescu,~I.~M.; Ashhab,~S.; Nori,~F. Quantum simulation. \emph{Rev.\ Mod.\ Phys.} \textbf{2014}, \emph{86}, 153\relax
\mciteBstWouldAddEndPuncttrue
\mciteSetBstMidEndSepPunct{\mcitedefaultmidpunct}
{\mcitedefaultendpunct}{\mcitedefaultseppunct}\relax
\EndOfBibitem
\bibitem[Berry \latin{et~al.}(2007)Berry, Ahokas, Cleve, and Sanders]{berry2007efficient}
Berry,~D.~W.; Ahokas,~G.; Cleve,~R.; Sanders,~B.~C. Efficient quantum algorithms for simulating sparse Hamiltonians. \emph{Commun. Math Phys.} \textbf{2007}, \emph{270}, 359--371\relax
\mciteBstWouldAddEndPuncttrue
\mciteSetBstMidEndSepPunct{\mcitedefaultmidpunct}
{\mcitedefaultendpunct}{\mcitedefaultseppunct}\relax
\EndOfBibitem
\bibitem[Hatano and Suzuki(2005)Hatano, and Suzuki]{hatano2005finding}
Hatano,~N.; Suzuki,~M. Finding Exponential Product Formulas of Higher Orders. \emph{Lect. Notes Phys.} \textbf{2005}, \emph{679}\relax
\mciteBstWouldAddEndPuncttrue
\mciteSetBstMidEndSepPunct{\mcitedefaultmidpunct}
{\mcitedefaultendpunct}{\mcitedefaultseppunct}\relax
\EndOfBibitem
\bibitem[Childs and Wiebe(2012)Childs, and Wiebe]{childs2012hamiltonian}
Childs,~A.~M.; Wiebe,~N. Hamiltonian simulation using linear combinations of unitary operations. \emph{arXiv preprint arXiv:1202.5822} \textbf{2012}, \relax
\mciteBstWouldAddEndPunctfalse
\mciteSetBstMidEndSepPunct{\mcitedefaultmidpunct}
{}{\mcitedefaultseppunct}\relax
\EndOfBibitem
\bibitem[Berry \latin{et~al.}(2015)Berry, Childs, Cleve, Kothari, and Somma]{BerChiCle15}
Berry,~D.~W.; Childs,~A.~M.; Cleve,~R.; Kothari,~R.; Somma,~R.~D. Simulating Hamiltonian Dynamics with a Truncated Taylor Series. \emph{Phys. Rev. Lett.} \textbf{2015}, \emph{114}, 090502\relax
\mciteBstWouldAddEndPuncttrue
\mciteSetBstMidEndSepPunct{\mcitedefaultmidpunct}
{\mcitedefaultendpunct}{\mcitedefaultseppunct}\relax
\EndOfBibitem
\bibitem[Lovett \latin{et~al.}(2010)Lovett, Cooper, Everitt, Trevers, and Kendon]{lovett2010universal}
Lovett,~N.~B.; Cooper,~S.; Everitt,~M.; Trevers,~M.; Kendon,~V. Universal quantum computation using the discrete-time quantum walk. \emph{Phys.\ Rev.~A} \textbf{2010}, \emph{81}, 042330\relax
\mciteBstWouldAddEndPuncttrue
\mciteSetBstMidEndSepPunct{\mcitedefaultmidpunct}
{\mcitedefaultendpunct}{\mcitedefaultseppunct}\relax
\EndOfBibitem
\bibitem[Somma(2016)]{somma2016trotter}
Somma,~R.~D. A Trotter-Suzuki approximation for Lie groups with applications to Hamiltonian simulation. \emph{J. Math Anal. Appl.} \textbf{2016}, \emph{57}\relax
\mciteBstWouldAddEndPuncttrue
\mciteSetBstMidEndSepPunct{\mcitedefaultmidpunct}
{\mcitedefaultendpunct}{\mcitedefaultseppunct}\relax
\EndOfBibitem
\bibitem[Low and Chuang(2017)Low, and Chuang]{low2017optimal}
Low,~G.~H.; Chuang,~I.~L. Optimal Hamiltonian simulation by quantum signal processing. \emph{Phys.\ Rev.\ Lett.} \textbf{2017}, \emph{118}, 010501\relax
\mciteBstWouldAddEndPuncttrue
\mciteSetBstMidEndSepPunct{\mcitedefaultmidpunct}
{\mcitedefaultendpunct}{\mcitedefaultseppunct}\relax
\EndOfBibitem
\bibitem[Kikuchi \latin{et~al.}(2023)Kikuchi, Mc~Keever, Coopmans, Lubasch, and Benedetti]{kikuchi2023realization}
Kikuchi,~Y.; Mc~Keever,~C.; Coopmans,~L.; Lubasch,~M.; Benedetti,~M. Realization of quantum signal processing on a noisy quantum computer. \emph{npj\ Quantum\ Inf.} \textbf{2023}, \emph{9}, 93\relax
\mciteBstWouldAddEndPuncttrue
\mciteSetBstMidEndSepPunct{\mcitedefaultmidpunct}
{\mcitedefaultendpunct}{\mcitedefaultseppunct}\relax
\EndOfBibitem
\bibitem[Childs \latin{et~al.}(2021)Childs, Su, Tran, Wiebe, and Zhu]{ChiSuTra21}
Childs,~A.~M.; Su,~Y.; Tran,~M.~C.; Wiebe,~N.; Zhu,~S. Theory of Trotter Error with Commutator Scaling. \emph{Phys. Rev. X} \textbf{2021}, \emph{11}, 011020\relax
\mciteBstWouldAddEndPuncttrue
\mciteSetBstMidEndSepPunct{\mcitedefaultmidpunct}
{\mcitedefaultendpunct}{\mcitedefaultseppunct}\relax
\EndOfBibitem
\bibitem[Preskill(2018)]{preskill2018quantum}
Preskill,~J. Quantum computing in the NISQ era and beyond. \emph{Quantum} \textbf{2018}, \emph{2}, 79\relax
\mciteBstWouldAddEndPuncttrue
\mciteSetBstMidEndSepPunct{\mcitedefaultmidpunct}
{\mcitedefaultendpunct}{\mcitedefaultseppunct}\relax
\EndOfBibitem
\bibitem[Tran \latin{et~al.}(2021)Tran, Su, Carney, and Taylor]{tran2021faster}
Tran,~M.~C.; Su,~Y.; Carney,~D.; Taylor,~J.~M. Faster digital quantum simulation by symmetry protection. \emph{PRX Quantum} \textbf{2021}, \emph{2}, 010323\relax
\mciteBstWouldAddEndPuncttrue
\mciteSetBstMidEndSepPunct{\mcitedefaultmidpunct}
{\mcitedefaultendpunct}{\mcitedefaultseppunct}\relax
\EndOfBibitem
\bibitem[Ward \latin{et~al.}(2009)Ward, Kassal, and Aspuru-Guzik]{ward2009preparation}
Ward,~N.~J.; Kassal,~I.; Aspuru-Guzik,~A. Preparation of many-body states for quantum simulation. \emph{J.~Chem.\ Phys.} \textbf{2009}, \emph{130}\relax
\mciteBstWouldAddEndPuncttrue
\mciteSetBstMidEndSepPunct{\mcitedefaultmidpunct}
{\mcitedefaultendpunct}{\mcitedefaultseppunct}\relax
\EndOfBibitem
\bibitem[Camps \latin{et~al.}(2022)Camps, K\"okc\"u, Bassman~Oftelie, De~Jong, Kemper, and Van~Beeumen]{camps2022algebraic}
Camps,~D.; K\"okc\"u,~E.; Bassman~Oftelie,~L.; De~Jong,~W.~A.; Kemper,~A.~F.; Van~Beeumen,~R. An algebraic quantum circuit compression algorithm for hamiltonian simulation. \emph{Siam. J. Matrix Anal. A} \textbf{2022}, \emph{43}, 1084--1108\relax
\mciteBstWouldAddEndPuncttrue
\mciteSetBstMidEndSepPunct{\mcitedefaultmidpunct}
{\mcitedefaultendpunct}{\mcitedefaultseppunct}\relax
\EndOfBibitem
\bibitem[Endo \latin{et~al.}(2020)Endo, Sun, Li, Benjamin, and Yuan]{endo2020variational}
Endo,~S.; Sun,~J.; Li,~Y.; Benjamin,~S.~C.; Yuan,~X. Variational quantum simulation of general processes. \emph{Phys.\ Rev.\ Lett.} \textbf{2020}, \emph{125}, 010501\relax
\mciteBstWouldAddEndPuncttrue
\mciteSetBstMidEndSepPunct{\mcitedefaultmidpunct}
{\mcitedefaultendpunct}{\mcitedefaultseppunct}\relax
\EndOfBibitem
\bibitem[Cerezo \latin{et~al.}(2021)Cerezo, Arrasmith, Babbush, Benjamin, Endo, Fujii, McClean, Mitarai, Yuan, Cincio, \latin{et~al.} others]{cerezo2021variational}
Cerezo,~M.; Arrasmith,~A.; Babbush,~R.; Benjamin,~S.~C.; Endo,~S.; Fujii,~K.; McClean,~J.~R.; Mitarai,~K.; Yuan,~X.; Cincio,~L.; others Variational quantum algorithms. \emph{Nat. Rev. Phys.} \textbf{2021}, \emph{3}, 625--644\relax
\mciteBstWouldAddEndPuncttrue
\mciteSetBstMidEndSepPunct{\mcitedefaultmidpunct}
{\mcitedefaultendpunct}{\mcitedefaultseppunct}\relax
\EndOfBibitem
\bibitem[Barison \latin{et~al.}(2021)Barison, Vicentini, and Carleo]{barison2021efficient}
Barison,~S.; Vicentini,~F.; Carleo,~G. An efficient quantum algorithm for the time evolution of parameterized circuits. \emph{Quantum} \textbf{2021}, \emph{5}, 512\relax
\mciteBstWouldAddEndPuncttrue
\mciteSetBstMidEndSepPunct{\mcitedefaultmidpunct}
{\mcitedefaultendpunct}{\mcitedefaultseppunct}\relax
\EndOfBibitem
\bibitem[Cirstoiu \latin{et~al.}(2020)Cirstoiu, Holmes, Iosue, Cincio, Coles, and Sornborger]{cirstoiu2020variational}
Cirstoiu,~C.; Holmes,~Z.; Iosue,~J.; Cincio,~L.; Coles,~P.~J.; Sornborger,~A. Variational fast forwarding for quantum simulation beyond the coherence time. \emph{npj\ Quantum\ Inf.} \textbf{2020}, \emph{6}, 82\relax
\mciteBstWouldAddEndPuncttrue
\mciteSetBstMidEndSepPunct{\mcitedefaultmidpunct}
{\mcitedefaultendpunct}{\mcitedefaultseppunct}\relax
\EndOfBibitem
\bibitem[Benedetti \latin{et~al.}(2021)Benedetti, Fiorentini, and Lubasch]{benedetti2021hardware}
Benedetti,~M.; Fiorentini,~M.; Lubasch,~M. Hardware-efficient variational quantum algorithms for time evolution. \emph{Phys.\ rev.\ res} \textbf{2021}, \emph{3}, 033083\relax
\mciteBstWouldAddEndPuncttrue
\mciteSetBstMidEndSepPunct{\mcitedefaultmidpunct}
{\mcitedefaultendpunct}{\mcitedefaultseppunct}\relax
\EndOfBibitem
\bibitem[K\"okc\"u \latin{et~al.}(2022)K\"okc\"u, Steckmann, Wang, Freericks, Dumitrescu, and Kemper]{KokSteWan22}
K\"okc\"u,~E.; Steckmann,~T.; Wang,~Y.; Freericks,~J.~K.; Dumitrescu,~E.~F.; Kemper,~A.~F. Fixed Depth Hamiltonian Simulation via Cartan Decomposition. \emph{Phys. Rev. Lett.} \textbf{2022}, \emph{129}, 070501\relax
\mciteBstWouldAddEndPuncttrue
\mciteSetBstMidEndSepPunct{\mcitedefaultmidpunct}
{\mcitedefaultendpunct}{\mcitedefaultseppunct}\relax
\EndOfBibitem
\bibitem[Khaneja and Glaser(2001)Khaneja, and Glaser]{khaneja2001cartan}
Khaneja,~N.; Glaser,~S.~J. Cartan decomposition of SU (2n) and control of spin systems. \emph{Chem.\ Phys.} \textbf{2001}, \emph{267}, 11--23\relax
\mciteBstWouldAddEndPuncttrue
\mciteSetBstMidEndSepPunct{\mcitedefaultmidpunct}
{\mcitedefaultendpunct}{\mcitedefaultseppunct}\relax
\EndOfBibitem
\bibitem[Mc~Keever and Lubasch(2024)Mc~Keever, and Lubasch]{mc2024towards}
Mc~Keever,~C.; Lubasch,~M. Towards adiabatic quantum computing using compressed quantum circuits. \emph{PRX Quantum} \textbf{2024}, \emph{5}, 020362\relax
\mciteBstWouldAddEndPuncttrue
\mciteSetBstMidEndSepPunct{\mcitedefaultmidpunct}
{\mcitedefaultendpunct}{\mcitedefaultseppunct}\relax
\EndOfBibitem
\bibitem[Bravyi \latin{et~al.}(2019)Bravyi, Browne, Calpin, Campbell, Gosset, and Howard]{bravyi2019simulation}
Bravyi,~S.; Browne,~D.; Calpin,~P.; Campbell,~E.; Gosset,~D.; Howard,~M. Simulation of quantum circuits by low-rank stabilizer decompositions. \emph{Quantum} \textbf{2019}, \emph{3}, 181\relax
\mciteBstWouldAddEndPuncttrue
\mciteSetBstMidEndSepPunct{\mcitedefaultmidpunct}
{\mcitedefaultendpunct}{\mcitedefaultseppunct}\relax
\EndOfBibitem
\bibitem[Hai and Ho(2023)Hai, and Ho]{hai2023universal}
Hai,~V.~T.; Ho,~L.~B. Universal compilation for quantum state tomography. \emph{Sci. Rep.} \textbf{2023}, \emph{13}, 3750\relax
\mciteBstWouldAddEndPuncttrue
\mciteSetBstMidEndSepPunct{\mcitedefaultmidpunct}
{\mcitedefaultendpunct}{\mcitedefaultseppunct}\relax
\EndOfBibitem
\bibitem[Jordan and Wigner(1928)Jordan, and Wigner]{jordan1928ber}
Jordan,~P.; Wigner,~E. $\ddot{\rm{U}}$ber das Paulische $\ddot{\rm{A}}$quivalenzverbot. \emph{Z. Phys.} \textbf{1928}, \emph{47}, 631--651\relax
\mciteBstWouldAddEndPuncttrue
\mciteSetBstMidEndSepPunct{\mcitedefaultmidpunct}
{\mcitedefaultendpunct}{\mcitedefaultseppunct}\relax
\EndOfBibitem
\bibitem[Bravyi and Kitaev(2002)Bravyi, and Kitaev]{bravyi2002fermionic}
Bravyi,~S.~B.; Kitaev,~A.~Y. Fermionic quantum computation. \emph{Ann. Phys.} \textbf{2002}, \emph{298}, 210--226\relax
\mciteBstWouldAddEndPuncttrue
\mciteSetBstMidEndSepPunct{\mcitedefaultmidpunct}
{\mcitedefaultendpunct}{\mcitedefaultseppunct}\relax
\EndOfBibitem
\bibitem[Peruzzo \latin{et~al.}(2014)Peruzzo, McClean, Shadbolt, Yung, Zhou, Love, Aspuru-Guzik, and O'~Brien]{PerMcCSha14}
Peruzzo,~A.; McClean,~J.; Shadbolt,~P.; Yung,~M.-H.; Zhou,~X.-Q.; Love,~P.~J.; Aspuru-Guzik,~A.; O'~Brien,~J.~L. A variational eigenvalue solver on a photonic quantum processor. \emph{Nat. Commun.} \textbf{2014}, \emph{5}, 4213\relax
\mciteBstWouldAddEndPuncttrue
\mciteSetBstMidEndSepPunct{\mcitedefaultmidpunct}
{\mcitedefaultendpunct}{\mcitedefaultseppunct}\relax
\EndOfBibitem
\bibitem[McClean \latin{et~al.}(2016)McClean, Romero, Babbush, and Aspuru-Guzik]{McCRomBab16}
McClean,~J.~R.; Romero,~J.; Babbush,~R.; Aspuru-Guzik,~A. The theory of variational hybrid quantum-classical algorithms. \emph{New J. Phys.} \textbf{2016}, \emph{18}, 023023\relax
\mciteBstWouldAddEndPuncttrue
\mciteSetBstMidEndSepPunct{\mcitedefaultmidpunct}
{\mcitedefaultendpunct}{\mcitedefaultseppunct}\relax
\EndOfBibitem
\bibitem[Shen \latin{et~al.}(2017)Shen, Zhang, Zhang, Zhang, Yung, and Kim]{SheZhaZha17}
Shen,~Y.; Zhang,~X.; Zhang,~S.; Zhang,~J.-N.; Yung,~M.-H.; Kim,~K. Quantum implementation of the unitary coupled cluster for simulating molecular electronic structure. \emph{Phys.\ Rev.~A} \textbf{2017}, \emph{95}, 020501\relax
\mciteBstWouldAddEndPuncttrue
\mciteSetBstMidEndSepPunct{\mcitedefaultmidpunct}
{\mcitedefaultendpunct}{\mcitedefaultseppunct}\relax
\EndOfBibitem
\bibitem[Romero \latin{et~al.}(2018)Romero, Babbush, McClean, Hempel, Love, and Aspuru-Guzik]{RomBabMcC18}
Romero,~J.; Babbush,~R.; McClean,~J.~R.; Hempel,~C.; Love,~P.~J.; Aspuru-Guzik,~A. Strategies for quantum computing molecular energies using the unitary coupled cluster ansatz. \emph{Quantum Sci. Technol.} \textbf{2018}, \emph{4}, 014008\relax
\mciteBstWouldAddEndPuncttrue
\mciteSetBstMidEndSepPunct{\mcitedefaultmidpunct}
{\mcitedefaultendpunct}{\mcitedefaultseppunct}\relax
\EndOfBibitem
\bibitem[Lee \latin{et~al.}(2019)Lee, Huggins, Head-Gordon, and Whaley]{LeeHugHea19}
Lee,~J.; Huggins,~W.~J.; Head-Gordon,~M.; Whaley,~K.~B. Generalized Unitary Coupled Cluster Wave functions for Quantum Computation. \emph{J.~Chem.\ Theory Comput.} \textbf{2019}, \emph{15}, 311--324\relax
\mciteBstWouldAddEndPuncttrue
\mciteSetBstMidEndSepPunct{\mcitedefaultmidpunct}
{\mcitedefaultendpunct}{\mcitedefaultseppunct}\relax
\EndOfBibitem
\bibitem[Kandala \latin{et~al.}(2017)Kandala, Mezzacapo, Temme, Takita, Brink, Chow, and Gambetta]{KanMezTem17}
Kandala,~A.; Mezzacapo,~A.; Temme,~K.; Takita,~M.; Brink,~M.; Chow,~J.~M.; Gambetta,~J.~M. Hardware-efficient variational quantum eigensolver for small molecules and quantum magnets. \emph{Nature} \textbf{2017}, \emph{549}, 242\relax
\mciteBstWouldAddEndPuncttrue
\mciteSetBstMidEndSepPunct{\mcitedefaultmidpunct}
{\mcitedefaultendpunct}{\mcitedefaultseppunct}\relax
\EndOfBibitem
\bibitem[Barkoutsos \latin{et~al.}(2018)Barkoutsos, Gonthier, Sokolov, Moll, Salis, Fuhrer, Ganzhorn, Egger, Troyer, Mezzacapo, Filipp, and Tavernelli]{BarGonSok18}
Barkoutsos,~P.~K.; Gonthier,~J.~F.; Sokolov,~I.; Moll,~N.; Salis,~G.; Fuhrer,~A.; Ganzhorn,~M.; Egger,~D.~J.; Troyer,~M.; Mezzacapo,~A.; Filipp,~S.; Tavernelli,~I. Quantum algorithms for electronic structure calculations: Particle-hole Hamiltonian and optimized wave-function expansions. \emph{Phys. Rev. A} \textbf{2018}, \emph{98}, 022322\relax
\mciteBstWouldAddEndPuncttrue
\mciteSetBstMidEndSepPunct{\mcitedefaultmidpunct}
{\mcitedefaultendpunct}{\mcitedefaultseppunct}\relax
\EndOfBibitem
\bibitem[Zeng \latin{et~al.}(2023)Zeng, Fan, Liu, Li, and Yang]{ZenFanLiu23}
Zeng,~X.; Fan,~Y.; Liu,~J.; Li,~Z.; Yang,~J. Quantum Neural Network Inspired Hardware Adaptable Ansatz for Efficient Quantum Simulation of Chemical Systems. \emph{J.~Chem.\ Theory Comput.} \textbf{2023}, \emph{19}, 8587--8597\relax
\mciteBstWouldAddEndPuncttrue
\mciteSetBstMidEndSepPunct{\mcitedefaultmidpunct}
{\mcitedefaultendpunct}{\mcitedefaultseppunct}\relax
\EndOfBibitem
\bibitem[Wiersema \latin{et~al.}(2020)Wiersema, Zhou, de~Sereville, Carrasquilla, Kim, and Yuen]{wiersema2020exploring}
Wiersema,~R.; Zhou,~C.; de~Sereville,~Y.; Carrasquilla,~J.~F.; Kim,~Y.~B.; Yuen,~H. Exploring entanglement and optimization within the hamiltonian variational ansatz. \emph{PRX Quantum} \textbf{2020}, \emph{1}, 020319\relax
\mciteBstWouldAddEndPuncttrue
\mciteSetBstMidEndSepPunct{\mcitedefaultmidpunct}
{\mcitedefaultendpunct}{\mcitedefaultseppunct}\relax
\EndOfBibitem
\bibitem[Tilly \latin{et~al.}(2022)Tilly, Chen, Cao, Picozzi, Setia, Li, Grant, Wossnig, Rungger, Booth, and Tennyson]{TilCheCao22}
Tilly,~J.; Chen,~H.; Cao,~S.; Picozzi,~D.; Setia,~K.; Li,~Y.; Grant,~E.; Wossnig,~L.; Rungger,~I.; Booth,~G.~H.; Tennyson,~J. The Variational Quantum Eigensolver: A review of methods and best practices. \emph{Phys. Rep.} \textbf{2022}, \emph{986}, 1--128\relax
\mciteBstWouldAddEndPuncttrue
\mciteSetBstMidEndSepPunct{\mcitedefaultmidpunct}
{\mcitedefaultendpunct}{\mcitedefaultseppunct}\relax
\EndOfBibitem
\bibitem[McLachlan(1964)]{mclachlan1964variational}
McLachlan,~A. A variational solution of the time-dependent Schrodinger equation. \emph{Mol. Phys.} \textbf{1964}, \emph{8}, 39--44\relax
\mciteBstWouldAddEndPuncttrue
\mciteSetBstMidEndSepPunct{\mcitedefaultmidpunct}
{\mcitedefaultendpunct}{\mcitedefaultseppunct}\relax
\EndOfBibitem
\bibitem[Cleve \latin{et~al.}(1998)Cleve, Ekert, Macchiavello, and Mosca]{cleve1998quantum}
Cleve,~R.; Ekert,~A.; Macchiavello,~C.; Mosca,~M. Quantum algorithms revisited. \emph{P.\ Roy.\ Soc.\ A-math.\ Phy} \textbf{1998}, \emph{454}, 339--354\relax
\mciteBstWouldAddEndPuncttrue
\mciteSetBstMidEndSepPunct{\mcitedefaultmidpunct}
{\mcitedefaultendpunct}{\mcitedefaultseppunct}\relax
\EndOfBibitem
\bibitem[Yuan \latin{et~al.}(2019)Yuan, Endo, Zhao, Li, and Benjamin]{yuan2019theory}
Yuan,~X.; Endo,~S.; Zhao,~Q.; Li,~Y.; Benjamin,~S.~C. Theory of variational quantum simulation. \emph{Quantum} \textbf{2019}, \emph{3}, 191\relax
\mciteBstWouldAddEndPuncttrue
\mciteSetBstMidEndSepPunct{\mcitedefaultmidpunct}
{\mcitedefaultendpunct}{\mcitedefaultseppunct}\relax
\EndOfBibitem
\bibitem[Earp and Pachos(2005)Earp, and Pachos]{EarPac05}
Earp,~H. N.~S.; Pachos,~J.~K. A constructive algorithm for the Cartan decomposition of SU(2N). \emph{J. Math. Phys.} \textbf{2005}, \emph{46}, 082108\relax
\mciteBstWouldAddEndPuncttrue
\mciteSetBstMidEndSepPunct{\mcitedefaultmidpunct}
{\mcitedefaultendpunct}{\mcitedefaultseppunct}\relax
\EndOfBibitem
\bibitem[Sun \latin{et~al.}(2018)Sun, Berkelbach, Blunt, Booth, Guo, Li, Liu, McClain, Sayfutyarova, Sharma, \latin{et~al.} others]{ns2018booth}
Sun,~Q.; Berkelbach,~T.~C.; Blunt,~N.~S.; Booth,~G.~H.; Guo,~S.; Li,~Z.; Liu,~J.; McClain,~J.~D.; Sayfutyarova,~E.~R.; Sharma,~S.; others PySCF: the Python-based simulations of chemistry framework. \emph{Wires. Comput. Mol. Sci.} \textbf{2018}, \emph{8}, e1340\relax
\mciteBstWouldAddEndPuncttrue
\mciteSetBstMidEndSepPunct{\mcitedefaultmidpunct}
{\mcitedefaultendpunct}{\mcitedefaultseppunct}\relax
\EndOfBibitem
\bibitem[Fan \latin{et~al.}(2022)Fan, Liu, Zeng, Xu, Shang, Li, and Yang]{q2chemistry}
Fan,~Y.; Liu,~J.; Zeng,~X.; Xu,~Z.; Shang,~H.; Li,~Z.; Yang,~J. $Q^{2}$ Chemistry: A quantum computation platform for quantum chemistry. \emph{arXiv preprint arXiv:2208.10978} \textbf{2022}, \relax
\mciteBstWouldAddEndPunctfalse
\mciteSetBstMidEndSepPunct{\mcitedefaultmidpunct}
{}{\mcitedefaultseppunct}\relax
\EndOfBibitem
\bibitem[Sun \latin{et~al.}(2022)Sun, Endo, Lin, Hayden, Vedral, and Yuan]{SunEndLin21}
Sun,~J.; Endo,~S.; Lin,~H.; Hayden,~P.; Vedral,~V.; Yuan,~X. Perturbative quantum simulation. \emph{Phys.\ Rev.\ Lett.} \textbf{2022}, \emph{129}, 120505\relax
\mciteBstWouldAddEndPuncttrue
\mciteSetBstMidEndSepPunct{\mcitedefaultmidpunct}
{\mcitedefaultendpunct}{\mcitedefaultseppunct}\relax
\EndOfBibitem
\bibitem[Zhang \latin{et~al.}(2023)Zhang, Sun, Yuan, and Yung]{ZhaSunYua23}
Zhang,~Z.-J.; Sun,~J.; Yuan,~X.; Yung,~M.-H. Low-Depth Hamiltonian Simulation by an Adaptive Product Formula. \emph{Phys. Rev. Lett.} \textbf{2023}, \emph{130}, 040601\relax
\mciteBstWouldAddEndPuncttrue
\mciteSetBstMidEndSepPunct{\mcitedefaultmidpunct}
{\mcitedefaultendpunct}{\mcitedefaultseppunct}\relax
\EndOfBibitem
\bibitem[Anthony(2006)]{anthony2006functionalized}
Anthony,~J.~E. Functionalized acenes and heteroacenes for organic electronics. \emph{Chem.\ Rev.} \textbf{2006}, \emph{106}, 5028--5048\relax
\mciteBstWouldAddEndPuncttrue
\mciteSetBstMidEndSepPunct{\mcitedefaultmidpunct}
{\mcitedefaultendpunct}{\mcitedefaultseppunct}\relax
\EndOfBibitem
\bibitem[Huang \latin{et~al.}(2022)Huang, Cai, Li, Ge, Hou, Li, Liu, Shi, Chen, Zheng, Xu, Liu, Li, Fan, and Fang]{HuaCaiLi22}
Huang,~K.; Cai,~X.; Li,~H.; Ge,~Z.-Y.; Hou,~R.; Li,~H.; Liu,~T.; Shi,~Y.; Chen,~C.; Zheng,~D.; Xu,~K.; Liu,~Z.-B.; Li,~Z.; Fan,~H.; Fang,~W.-H. Variational Quantum Computation of Molecular Linear Response Properties on a Superconducting Quantum Processor. \emph{J.~Phys.\ Chem.\ Lett.} \textbf{2022}, \emph{13}, 9114--9121\relax
\mciteBstWouldAddEndPuncttrue
\mciteSetBstMidEndSepPunct{\mcitedefaultmidpunct}
{\mcitedefaultendpunct}{\mcitedefaultseppunct}\relax
\EndOfBibitem
\bibitem[Clar and Schoental(1964)Clar, and Schoental]{clar1964polycyclic}
Clar,~E.; Schoental,~R. \emph{Polycyclic hydrocarbons}; Springer, 1964; Vol.~2\relax
\mciteBstWouldAddEndPuncttrue
\mciteSetBstMidEndSepPunct{\mcitedefaultmidpunct}
{\mcitedefaultendpunct}{\mcitedefaultseppunct}\relax
\EndOfBibitem
\bibitem[Pedernales \latin{et~al.}(2014)Pedernales, Di~Candia, Egusquiza, Casanova, and Solano]{pedernales2014efficient}
Pedernales,~J.; Di~Candia,~R.; Egusquiza,~I.; Casanova,~J.; Solano,~E. Efficient quantum algorithm for computing n-time correlation functions. \emph{Phys.\ Rev.\ Lett.} \textbf{2014}, \emph{113}, 020505\relax
\mciteBstWouldAddEndPuncttrue
\mciteSetBstMidEndSepPunct{\mcitedefaultmidpunct}
{\mcitedefaultendpunct}{\mcitedefaultseppunct}\relax
\EndOfBibitem
\bibitem[Francis \latin{et~al.}(2020)Francis, Freericks, and Kemper]{FraFreKem20}
Francis,~A.; Freericks,~J.~K.; Kemper,~A.~F. Quantum computation of magnon spectra. \emph{Phys. Rev. B} \textbf{2020}, \emph{101}, 014411\relax
\mciteBstWouldAddEndPuncttrue
\mciteSetBstMidEndSepPunct{\mcitedefaultmidpunct}
{\mcitedefaultendpunct}{\mcitedefaultseppunct}\relax
\EndOfBibitem
\bibitem[Fetter and Walecka(2012)Fetter, and Walecka]{fetter2012quantum}
Fetter,~A.~L.; Walecka,~J.~D. \emph{Quantum theory of many-particle systems}; Courier Corporation, 2012\relax
\mciteBstWouldAddEndPuncttrue
\mciteSetBstMidEndSepPunct{\mcitedefaultmidpunct}
{\mcitedefaultendpunct}{\mcitedefaultseppunct}\relax
\EndOfBibitem
\end{mcitethebibliography}
\bibliographystyle{achemso}
}

\end{document}